\documentclass[twocolumn, twocolappendix]{aastex631}

\newcommand*\chem[1]{\ensuremath{\mathrm{#1}}}

\begin{document}

\title{An Agnostic Biosignature Based on Modeling Panspermia and Terraforming}

\author[0000-0001-8806-9659]{Harrison B. Smith}
\altaffiliation{hbs@elsi.jp — HBS conceived original project idea, wrote and ran simulation code, wrote and edited the manuscript, and contributed to figure generation and data analysis}
\affiliation{Earth-Life Science Institute \\ Institute of Science Tokyo \\ Ookayama, Meguro-ku, Tokyo, Japan}
\affiliation{Blue Marble Space Institute of Science \\ Seattle, Washington, USA}

\author[0000-0003-2270-2954]{Lana Sinapayen}
\altaffiliation{lana.sinapayen@gmail.com — LS contributed to the experimental design, analysed the data, produced the figures, and edited the manuscript.
\\
\\
The code necessary to run the simulations and analyses is available at \url{https://github.com/hbsmith/SmithSinapayen2024}.
}
\affiliation{Sony Computer Science Laboratories, Kyoto, Japan}
\affiliation{National Institute for Basic Biology, Okazaki, Japan}




\begin{abstract}

    The search for a second instance of life is one of the greatest problems of modern science. Outside of creating an artificial origin of life on Earth, the primary targets for the search for life are planets inside or outside the solar system. Realistically, there are just a few locations to search for alien life within the solar system. Outside the solar system, opportunities are nearly unlimited, but there’s a catch: it is difficult to attribute, with certainty, features of exoplanets to extraterrestrial life. Simple spectral biosignatures are susceptible to false positives; technosignatures reduce this susceptibility at the expense of strong assumptions about potential underlying life and its technologies.  We have developed an agnostic approach to exoplanet life detection that overcomes these limitations by using properties that emerge on the scale of groups of planets, without the need for a “smoking-gun” single-planet level biosignature. We use an agent-based model to show that if life can spread between star systems, and affect the observable properties of a planet, then a robust signature of life (with very few false positives) can emerge, defined by correlations between planet characteristics and their locations. By clustering planets based only on their observed characteristics, and retaining clusters localized in space, we demonstrate (and evaluate) a way to prioritize specific planets for further observation, based on their potential for containing life. We consider obstacles that must be overcome to practically implement our approach, including identifying specific ways in which better understanding astrophysical and planetary processes would improve our ability to detect life.

\end{abstract}




\section{Introduction} \label{sec:intro}

It is difficult to attribute, with certainty, observable features of exoplanets to extraterrestrial life  \citep{moore_how_2017, tasker_language_2017, green_call_2021, cockell_standards_2022, lenardic_hype_2022, lenardic_communicating_2023, smith_life_2023}. This is exemplified by the insufficient (and often ill-defined) working definitions of life that are used to interpret observational data, and by the growing number of false positives for traditional biosignatures \citep{cleland_defining_2002, benner_defining_2010, cleland_life_2012, mix_defending_2015, schwieterman_identifying_2016, bich_defining_2018, harman_biosignature_2018, schwieterman_exoplanet_2018, mariscal_life_2020, janin_exoplanet_2021, vickers_confidence_2023}. 

One way in which the community aims to overcome the problems with defining life is to develop “agnostic” biosignatures\footnote{While the phrase agnostic biosignatures is relatively new, the concept of designing non-Earth centric life detection experiments dates back to at least 1961 \citep{sagan_origin_1961, lovelock_physical_1965}.}—detecting signs of life that are not particular to Earth-life or any other hypothetical instances of life \citep{marshall_identifying_2021, smith_grayness_2021}.  Yet, agnostic biosignature proposals are sometimes built on restricted concepts of habitability (e.g., requiring rocky, watery planets), or simple anomaly detection (i.e., without using a working definition of life)  \citep{kinney_epistemology_2022, cleaves_robust_2023}. 

A proposed remedy for the issues surrounding traditional biosignatures has been to use so-called ``statistical" biosignatures—relying on integrating multiple lines of evidence to increase confidence in a discovery, or generating ensembles of data to better constrain the probability of making specific observations contingent on the presence of life \citep{lin_statistical_2015, catling_exoplanet_2018, walker_exoplanet_2018, affholder_bayesian_2021, bixel_bioverse_2021}. However, such statistical biosignatures often rely on assumptions about the prior probability of abiogenesis, or on the existence of unambiguous biosignatures for single planets. These approaches are often trying to solve more specific issues, like estimating the frequency of Earth-like life, the frequency of planets originating from panspermia compared to abiogenesis, or finding evidence of a specific metabolic process \citep{lin_statistical_2015, walker_exoplanet_2018, affholder_bayesian_2021, checlair_probing_2021, kovacevic_possible_2022}.

Here we ask, can we detect the presence of life if we postulate that life is spreading between and terraforming planets?\footnote{Here, ``life spreading" refers to interstellar panspermia, and ``terraforming planets" refers to modifying observable characteristics of planets.} This is an astrobiological ``hinge proposition"\footnote{An assumption that enables a specific method of inquiry but which the method itself cannot verify. As \citet{kinney_epistemology_2022} write, “...it may still be the case that such assumptions are needed in order for us to have any chance of conducting successful inquiry in astrobiology."} of the kind described by Kinney \& Kempes as necessary to make sure ``that the door of astrobiology can turn properly" under conditions of deep uncertainty \citep{kinney_epistemology_2022}.  It is also somewhat the converse of a question posed in other panspermia related work: Can the prevalence of panspermia be constrained when assuming we have a way to detect life? \citep{lin_statistical_2015, balbi_quantifying_2020, grimaldi_feasibility_2021, lingam_birthdeathmigration_2021, kovacevic_possible_2022}. 

In fact, our postulates (of terraforming and panspermia) are less peculiar than they might at first seem: while existing work doesn’t always explicitly refer to terraforming, it must be assumed—in the sense that the mere presence of life stable over geological timescales would create environmental feedback with a planet, and if that life is to be detectable it must modify a planet's observables. Such planetary modification by life is a well-documented phenomena—e.g., the rise of \chem{O_2} during the great oxygenation event \citep{olejarz_great_2021}, or the rise of \chem{CO_2} from human industrial activity \citep{lynas_greater_2021}. Both of these examples represent unintentional terraforming, but  intentional terraforming from intelligent life has been an area of active research \citep{beech_terraforming_2021,fogg_terraforming_1998,zubrin_technological_1993}, and would be expected to cause environmental change at a faster timescale. 

The feasibility of interstellar lithopanspermia (non-intelligent exchange of material via rocks) has been discussed at length in other work (see e.g. \citealt{carroll-nellenback_fermi_2019, grimaldi_feasibility_2021, gobat_panspermia_2021, totani_solid_2023, cao_implications_2024} and references therein). Crucially, it appears plausible, although likelihoods and rates of material exchange vary considerably depending on assumptions on timing, amount of material ejected, organismal hardiness, and capture rates, among other features (e.g., recent work estimates that lithopanspermia is a plausible life-seeding mechanism for $\approx 10^5 $ planets in our Galaxy, but this estimate includes only Earth-sized planets in the habitable zone, \citealt{cao_implications_2024}). Arguments have likewise been made for the feasibility of directed interstellar panspermia by intelligent life \citep{wright_infrared_2014, newman_galactic_1981, crick_directed_1973}. Ultimately, our postulates of panspermia and terraforming are merely well understood hallmarks of life (proliferation via replication, and adaptation with bi-directional environmental feedback), escalated to the planetary scale, and executed on an interstellar scale.

We use simulations to show that statistical correlations between the spatial distribution of planets around different host stars, and their observable characteristics \textit{could itself} be evidence of life, without the need for a separate biosignature that could reliably detect life on any given planet in isolation. The agnosticism of our biosignature is inseparable from its emergence at the scale of a population of planets—singleton planetary anomalies might be explained away by unknown geochemical processes, or targeted simply because they are anomalous (without a clear hypothesis of why they should be explained by life). Hypothesizing that life spreads via panspermia and terraforming allows us to search for biosignatures while forgoing any strong assumptions about not only the peculiarities of life (e.g., its metabolism) and planetary habitability (e.g., requiring surface liquid water, \citealt{gobat_panspermia_2021}), but even the potential breadth of structure and chemical complexity underpinning living systems \citep{sole_largescale_2004, kim_universal_2019, bartlett_assessing_2022, wong_networkbased_2023}. 

We first describe our model (Sec. \ref{subsec:model}) and approach for statistically identifying the presence of terraformed planets (Sec. \ref{subsec:biosignature}). We show that evolving our model in time can increase the correlation between planets’ compositions and positions (Sec. \ref{subsec:simplecorr}). To identify specific planets likely to have been terraformed, we cluster planets by observable characteristics, then select clusters which are spatially localized and cause a decrease in correlation when removed (Secs. \ref{subsubsec:cluster}, \ref{subsubsec:selection}). We evaluate these clusters by calculating how well they correctly classify terraformed planets as being terraformed, and non-terraformed planets as being non-terraformed (Sec. \ref{subsec:promising}). Finally, we discuss how our results might change due to theoretical and observational constraints, identify specific ways in which better understanding astrophysical and planetary processes could improve life detection, and speculate on the concept of life at the scale of populations of planets.



\section{Methods} \label{sec:methods}
\subsection{Modeling Panspermia and Terraforming}\label{subsec:model}
\subsubsection{Summary}\label{subsubsec:model_summary}
\begin{figure}[b!]
\plotone{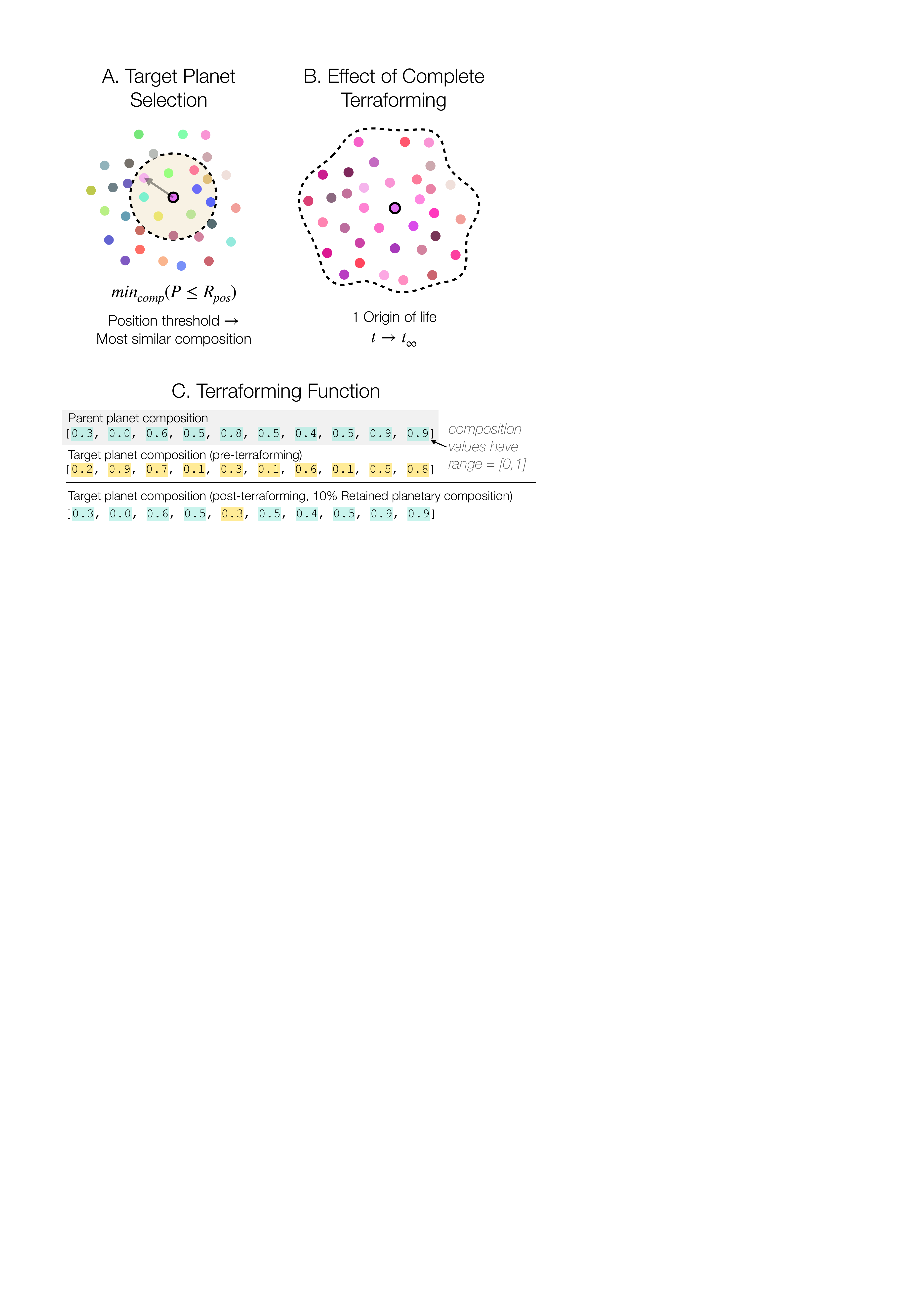}
\caption{
Target planet selection and terraforming. A. The objective function, used for determining the destination of life from a terraformed ``parent" planet. Candidate destinations are first constrained by a maximum positional distance threshold; among these candidates, the planet closest in composition to the parent planet is chosen as the target. B. Simulations are initialized with 1 origin of life, causing the initial distribution of planet compositions (seen in A) to become correlated. C. An example of how we determine target planet composition when retaining 10\% of the pre-terraformed planet composition. Note that while our simulations use a 3D space, the concept figure shows only  a 2D space for clarity.
\label{fig:concept}}
\end{figure}

We created an agent-based model to simulate life\footnote{In this work, we use \textit{life} specifically to refer to an agent (used strictly in the ``entity" sense of the word agent) traveling from a parent terraformed planet to a target non-terraformed planet, and whose arrival at the target planet triggers terraforming.} spreading between planets orbiting different host stars.  The model is initialized with 1000 planets (one of which is terraformed/panspermia-capable), fixed in a frame of reference, and uniformly randomly distributed in a continuous 3D volume. All planets are assumed to be in different star systems (i.e., we assume one planet per star). ``Planets" have compositions representing the observable characteristics of the planet, and ``life" agents have compositions representing the observable characteristics of the planet which the life originates from. During simulation, terraformed planets \footnote{We specifically avoid saying living planets, because our methodology cannot distinguish between living planets incapable of panspermia and non-living planets.} emit life towards a target non-terraformed planet at a constant velocity \footnote{This approach is most similar to directed panspermia by intelligent life, but could also be envisioned as undirected lithopanspermia, where we refrain from burdening the simulations with agents that won't ever be able to interact with planets under the rules of our model.}. The target planets are selected based on proximity in position and in composition (see Sec. \ref{subsubsec:target_selection}). When life arrives at a planet, the planet's composition is modified (terraformed) based on both the initial composition of the planet and the composition of the incoming life. Life originating from that terraformed planet will then reflect its new composition. The time that life takes to travel to a planet in our model can be thought of as the sum of the time it takes to travel to and terraform a planet. Terraformed planets emit life at a fixed rate, provided suitable target planets exist (see Sec.  \ref{subsubsec:target_selection}). The simulation ends when all planets are terraformed, or no suitable target planets remain (Fig. \ref{fig:concept}B).

\subsubsection{Compositions and Terraforming}  \label{subsubsec:compositions}
Both ``life" and ``planets" have compositions represented by vectors of 10 real numbers (each element $\in[0,1]$). In our model, these compositions are abstractions, used as a means to compare the compatibility of any given life and planet. While we might in the abstract think of one vector representing the observable characteristics of a planet, this same vector would not represent the \textit{same} observable characteristics of life compatible with that planet. For life, it instead represents the kind of \textit{planetary} observable characteristics which are compatible with that life. For example, perhaps a planet's composition vector $V$ corresponds to an atmosphere composed of 95\% \chem{CO_2} and 5\% \chem{N_2}. The same composition vector $V$ in the context of life is used to refer to a kind of life which could be compatible with a planet with an atmospheric composition of 95\% \chem{CO_2} and 5\% \chem{N_2}. This does not mean that life on such a planet is literally composed of 95\% \chem{CO_2} and 5\% \chem{N_2}. This captures the idea that if a life and a planet have identical composition vectors, then that life will surely be able to terraform the planet\footnote{This is an extreme example where technically life wouldn't even need to terraform the planet, because it implies that the planet's observable characteristics already perfectly match the planet from where the life originated.}. Correspondingly, in our model a life’s composition is identical to the planet from which it is emitted. Initial compositions of each planet are random, with each element in each composition vector drawn from a continuous uniform distribution $\in[0,1]$. This is a loose approximation of the unknowns surrounding the distribution of diversity of planets between planetary systems \citep{tasker_planetary_2020, pacetti_chemical_2022}.

Planet compositions are modified by terraforming, where a mixing function determines how to generate the terraformed composition based on the non-terraformed planet composition and the incoming life's composition. We apply a mixing function inspired by horizontal gene transfer \citep{senaratna_genetic_2005, umbarkar_crossover_2015}, but other mixing functions could also be used. The most important feature we aim to capture is the bi-directional feedback between the terraforming life, and the terraformed planet—such that future life emitted from the terraformed planet will inherit the post-terraforming characteristics of that planet. We keep  $n = 1$ randomly chosen element from the planet’s pre-terraformed composition, and the rest of the terraformed composition is inherited from the terraforming life (Fig. \ref{fig:concept}C). 

\subsubsection{Model Stepping}  \label{subsubsec:model_stepping}
Each step of the model increments the time by $dt = 10$. Time units are arbitrary, and only impact the distance that life travels in a step, $dt*v_{\text{life}}$. Here, $v_{\text{life}} = 0.2$, and $emit\_rate = 0.01$, meaning that for every 10 model steps, a planet can emit 1 new life, and existing life can travel 20 units (for reference, the average distance between planets is $\sim 20$ units).

\subsubsection{Target Planet Selection}  \label{subsubsec:target_selection}
The destination selection aims to balance the need of identifying a target planet that is both compositionally similar, and physically near, the parent planet. This is based on the assumption that life directing panspermia would want to minimize travel time and terraforming energy costs. In the context of lithopanspermia, this could be interpreted as an assumption that beyond a certain positional distance capture becomes vanishingly unlikely, and beyond a certain compositional distance compatibility becomes vanishingly unlikely. 

The algorithm we used for determining a life's target planet has a positional distance threshold, $R_{pos}$. Among all non-terraformed planets, $P$, under that positional distance threshold from the life's origin planet, it selects the planet nearest in compositional space, $min_{comp}(P\leq R_{pos})$. Here, $P \leq R_{pos}$, is shorthand for all non-terraformed planets within $R_{pos}$ of life's origin planet, and $min_{comp}(P)$ is the planet at minimum compositional distance to the life's origin planet (Fig. \ref{fig:concept}A). The distances calculated are Euclidean. A range of positional thresholds, $R_{pos}$, corresponding to $\approx 5\%, 11\%, 29\%,$ and $100\%$ of the maximum distance possible between any two points in space were chosen for $min_{comp}(P\leq R_{pos})$ in order to span the transition between a model where terraforming of every planet is impossible and every planet is possible, based on the average distance between planets. 

\subsection{Identifying the Presence of Terraformed Planets}\label{subsec:biosignature}
\subsubsection{Summary}\label{subsubsec:biosig_summary}
We hypothesize that the process of panspermia and terraforming in our model will lead to a population of planets with anomalously high positive correlations between their spatial locations and compositions, compared to random permutations of these planets' compositions. We quantify this using the Mantel test—a statistical test common in ecological science (Sec. \ref{subsubsec:mantel}). By clustering on the planet compositions, we can identify clusters of planets driving these correlations (Sec. \ref{subsubsec:cluster}). From these initial clusters, we select those localized in space (via their interquartile range, IQR), because we hypothesize that life would not only change the distribution of planetary compositions, but would also do so in a relatively compact portion of the galaxy. We further select clusters which, when removed, cause a decrease in the Mantel coefficient of the residual space of planets (Sec. \ref{subsubsec:selection}). Finally, we attempt to evaluate how well our clusters reflect the presence of truly terraformed planets (Sec. \ref{subsubsec:evaluation}).

\subsubsection{Mantel Test}\label{subsubsec:mantel}
The Mantel test\footnote{While originally developed for biostatistics, and often applied to ecological data, we use the Mantel test here because it is a straightforward and established way to measure the correlation between a distance matrix representing space, and one representing an arbitrary attribute. The Mantel test has both critics and advocates \citep{legendre_should_2015, somers_putting_2022}, which we mention simply to make clear that the approach we describe is not wedded to using the Mantel test—we would expect similar results when using other distance-based tests.} is a measure of correlation between two distance matrices. The resulting value is called the Mantel coefficient, and is reported alongside a p-value calculated from an approximate permutation test (indicating the proportion of randomly permuted distance matrices which have correlations greater or equal to the correlation between the non-permuted distance matrices). Simply put, the p-value indicates how unlikely this correlation is compared to random permutations of the data. We wrote the Mantel test in Julia, with the algorithm and code adapted from Python’s scikit-bio \citep{thescikit-biodevelopmentteam_scikitbio_2020, rideout_biocore_2023}. Here we used the Mantel test to measure the Pearson correlation between a distance matrix of all planet positions, and a distance matrix of all planet compositions. For each Mantel coefficient calculation, the p-value was generated using $99$ permutations with a 2-sided alternative hypothesis. This is approximately equal to $2.5\sigma$ confidence. Because the p-value quantifies how anomalous an observed composition/position association is, given the assumptions of the model that correlations should only occur from panspermia and terraforming, we treat the p-value as one measure of confidence in the space containing a biosignature. To get an idea of how sensitive the Mantel coefficient corresponding to a $2.5\sigma$ detection is, we plotted how it varies based on number of planets observed (Fig. \ref{fig:test_power}). We find that the sensitivity of the Mantel coefficient to number of planets observed decreases exponentially, and 1000 planets seems like a reasonable choice to reflect the balance of the challenge of realistically observing planets, with the need for those planets to exhibit potentially small correlations in composition-position space. The exact shape of this plot will vary by model parameters, but is especially dependent on the distribution of planet compositions and positions (e.g., planets being evenly distributed in composition or position space, vs. extremely heterogeneous).

\subsubsection{Clustering} \label{subsubsec:cluster}
We clustered planets in each iteration of the simulation based only on their compositions, using the DBSCAN algorithm implemented in R \citep{hahsler_dbscan_2019}. Briefly, compared to other clustering algorithms, DBSCAN does not require the number of clusters to be predefined, can identify clusters of varying shapes and sizes based on the density of data points in the feature space, and separates points as belonging to clusters or noise. Arguments were chosen based on advice in the documentation, with  $\texttt{minPts} = 11$ (the dimensionality of the data + 1), and $\texttt{eps}$ was chosen dynamically at each iteration based on the location of the elbow in the k-nearest neighbor distance plot. Because the location of the elbow is not always obvious, we used the R implementation of the Kneedle algorithm~\citep{satopaa_finding_2011}, with $\texttt{sensitivity parameter}=1$, empirically determined based on visual examination of the elbow placement on the nearest neighbor curves (Fig. \ref{fig:elbow}). Note that the sensitivity should be adjusted to the specific data being analyzed, and in our case the data changes at each time step. This results in the Kneedle sensitivity being appropriate for only part of the time steps in our simulation. We chose to focus on the early steps of the simulation (because the early steps of the simulation represent the time when life is not ubiquitous in the galaxy and thus expected to be harder to find), but the sensitivity should be adjusted if finding the curves' elbows in later time steps. DBSCAN classifies each point (planet) as either being part of a particular cluster, or noise (meaning it does not meet the criteria to fall into a cluster based on our chosen arguments). 

\subsubsection{Selecting clusters} \label{subsubsec:selection}
We first sought to select clusters for their likelihood of containing terraformed planets, without the aid of ground truth labels (i.e., without using our knowledge of which planets were truly terraformed in the model). We selected them by measuring their spatial spread, and by analyzing how the Mantel coefficient of the residual space of planets changes when removing clusters of planets.

We created a threshold for the spatial spread of the planets in the clusters by taking the average of the interquartile range (IQR, where the middle 50\% of planets fall) of all planets across each of the $x, y, z$ dimensions, $\frac{(\text{IQR}_x + \text{IQR}_y + \text{IQR}_z)}{3}$, within the spatial extent of a cube with sides of length 1/2 the size of the model space (i.e., with a volume = $1/8$ the model space). For the simulation analyzed in this paper, $\text{IQR}= 25.2$. Thus, any cluster with $\text{IQR} \leq 25.2$ meets our threshold for being spatially localized\footnote{Though the extent we chose for an IQR threshold is arbitrary, it reflects the relative size of our model space and the presumption that looking for life that is still relatively spatially localized is of greater interest than looking for life which has already spread over the galaxy (the latter implying there might be other easier ways to look for life). This choice, like others, could be modified depending on other assumptions or objectives. }. 

On top of this spatial IQR threshold, we imposed a second condition for selecting clusters: their impact on the full model space's Mantel coefficient when removed. We call the Mantel coefficient of the remaining space the residual mantel coefficient ($M_{residual}$). We quantify the impact of a cluster's removal relative to the original space's mantel coefficient ($M_{original}$), and call this the \textit{Mantel contribution} (MC), where $\text{MC} = \frac{M_{original}-M_{residual}}{M_{original}}$ . Our logic being that if $M_{residual}$ decreases with the exclusion of a subset of planets, those planets are likely a significant driver behind $M_{original}$, and thus would be favorable clusters for containing terraformed planets. We select any cluster with $\text{MC} \geq 0$ as meeting our MC threshold. 

\subsubsection{Evaluating clustering} \label{subsubsec:evaluation}
We evaluated selected clusters meeting our criteria ($IQR \leq 25.2 \land \text{MC} \geq 0$) for how well they identified terraformed/non-terraformed planets, based on the true labels of each planet throughout our simulation.  

At each iteration, for each selected cluster, we calculated the ratio of planets in the cluster which were terraformed (true positives, TP) and non-terraformed (false positives, FP), as well as the ratio of planets outside the cluster which were non-terraformed (true negatives, TN) and terraformed (false negatives, FN) (Fig. \ref{fig:perf_ratios}). For example, if 100 of 1000 planets are terraformed, and a single selected cluster is identified with 80 planets, only 75 of which are terraformed, then the ratios for each metric in this cluster are: $\text{TP} = 75/80$, $\text{FP} = 5/80$, $\text{TN}=895/920$, $\text{FN}=25/920$.

To simplify how this information is conveyed, we report the summary statistics of sensitivity, specificity, and accuracy.
\begin{itemize}
\item Sensitivity is the proportion of all terraformed planets correctly selected by the cluster, $\text{TP/(TP+FN)}$. This is $75/(75+25)$ in our example. 
\item Specificity is the proportion of all non-terraformed planets correctly \textit{not} selected by the cluster, $\text{TN/(TN+FP)}$. This is $895/(895+5)$ in our example.
\item Accuracy is the proportion of all planets correctly classified, $\text{TP+TN/(TP+FP+TN+FN)}$. This is $(75+895)/1000$ in our example. 
\end{itemize}
We believe a reliable biosignature must minimize false positives (i.e., must not misclassify non-terraformed planets), even at the expense of producing false negatives (i.e., missing terraformed planets). We thus consider our evaluation to be successful in validating our approach if specificity is high, even if sensitivity is low. 


\subsection{Software and Availability} \label{subsec:software}
The code necessary to run the simulations and analyses is available on Github at \url{https://github.com/hbsmith/SmithSinapayen2024}. Simulations were built in Julia, and analyses were carried out in Julia, Python, and R.


\section{Results} \label{sec:results}

\subsection{Panspermia can increase the correlation between planets’ compositions and positions} \label{subsec:simplecorr}

We find that panspermia causes uncorrelated planets to become positively correlated in position-composition distance space, as quantified by the Mantel test (see Methods. \ref{subsubsec:mantel}). 

\begin{figure}[ht!]

\epsscale{1.2}
\plotone{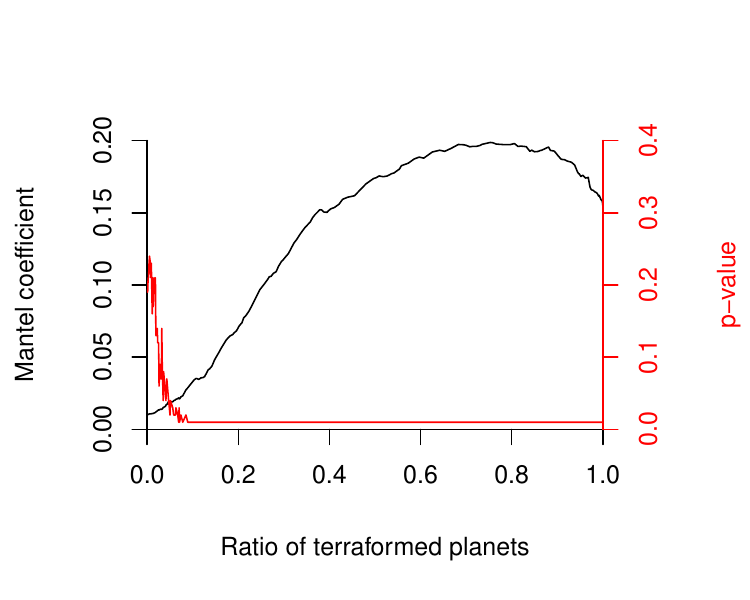}
\caption{
Mantel coefficient and p-value as a function of the ratio of planets terraformed. The earliest we observe a p-value $\leq 0.01$ is at a terraformed ratio $\approx 7\%$ (here, 70 planets).
\label{fig:mantel_pval}}
\end{figure}

We initialized a simulation with the intention of making terraforming easy and reliable, in order to check if a positive Mantel correlation coefficient is observable in a “best case” scenario. The simulation begins with a single terraformed planet, targets the most compositionally similar planet within $R_{pos} \leq 20$ ($\approx 11\%$ of the maximum possible distance between planets in our model), and with terraforming keeping 1 of the 10 compositional elements from the pre-terraformed planet. This means that over time, all planets in the simulation acquire similar, but not identical compositions. We found that the Mantel coefficient indeed increased as the ratio of terraformed planets increased, until reaching a peak around a terraformed ratio of $\approx 0.75$, and then decreased (Fig. \ref{fig:mantel_pval}, black). Because the Mantel coefficient measures the correlation between positional and compositional distance matrices of planets, this correlation will decrease if either distance matrix becomes too homogeneous. This is exactly what begins to happen with the compositional distance matrix, with the decline especially pronounced if there is perfect replication (see Appendix Fig. \ref{fig:param_sweep}, and Appendix Sec. \ref{sec:appendix} for a discussion on the effect of varying simulation parameters).

While the increase in Mantel correlation coefficient we can observe is striking, what we are really interested in is how anomalous the observed correlations in position/composition space are, and not the absolute value of them. That is, we expect life spreading via panspermia and terraforming to cause not just positive mantel correlation coefficients, but anomalously high ones compared to other possible permutations of the data. As seen in Fig. \ref{fig:mantel_pval} (in red), the p-value of our Mantel test indeed decreases as planets become terraformed, and after $\approx 8\%$ of planets are terraformed, $p \leq 0.01$ (which is the maximal precision we can reach with 99 permutations).

Based on our postulates, this indicates that the Mantel test—and more specifically its p-value (which remains low even after the coefficient decreases)—can identify the presence of panspermia and terraforming, provided there are enough planets in our observed population terraformed. When we investigated how sensitive the Mantel coefficient is to number of planets observed (independent of the number terraformed), we find that the value of the Mantel coefficient corresponding to a similarly confident $2.5\sigma$ detection increases exponentially with fewer planets observed (i.e., to reach the same confidence with fewer observations, we need a higher Mantel coefficient. See Fig. \ref{fig:test_power}). 

But even if we live at a point in time where there actually are $\geq 10\%$ of planets terraformed, this does not answer the question of \textit{which} planets observed within our population might host life. Identifying such planets would be important for determining targets for follow up detailed observations, in the event that the methodology adhered to with these model results could be carried out with a simpler set of observational data.

\subsection{Likely terraformed planets can be identified from clustering} \label{subsec:promising}

\begin{figure}[htb!]
\epsscale{1.2}
\plotone{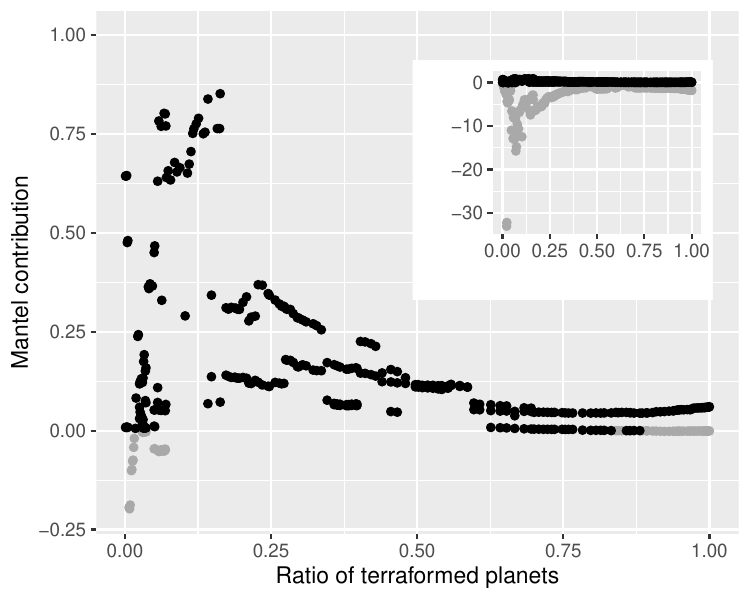}
\caption{
Mantel contribution as a cluster selection criterion. Clusters with a Mantel contribution $> 0$ (black) meet this selection criterion, indicating that their removal is a detriment to the residual space's Mantel coefficient (causing it to decrease). Clusters with a negative Mantel contribution shown in grey. The inset shows the full range of negative values in all clusters. 
\label{fig:mantel_contribution}}
\end{figure}

\begin{figure}[htb!]
\epsscale{1.2}
\plotone{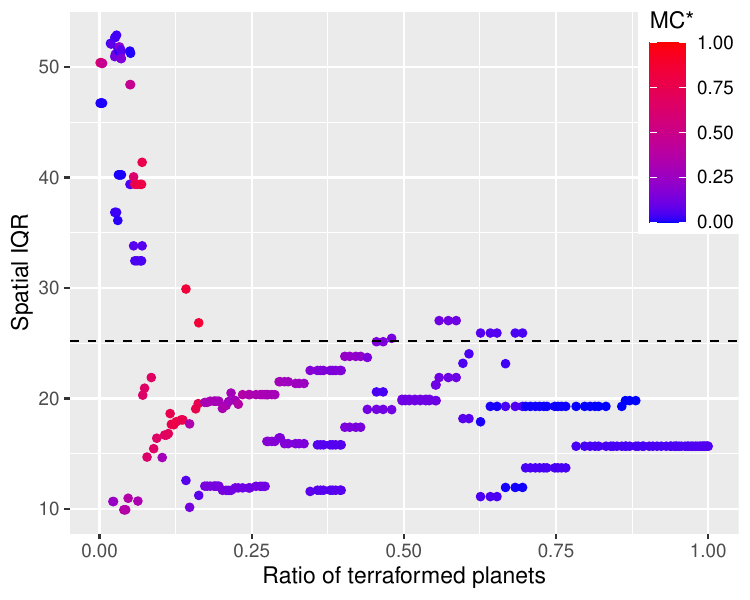}
\caption{
Selection criteria used on clusters of planets. Spatial localization of clusters of planets is shown on the y-axis, as measured by the Interquartile Range (IQR) of each cluster. Horizontal dashed line (at $\text{IQR}=25.2$) denotes the threshold used, below which we selected clusters for being spatially localized. This corresponds to approximately the average IQR of planets in a cube the size of 1/8 the model space. Color bar shows the Mantel Contribution (MC) of clusters, with a high MC indicating a cluster as being important for raising the full space's Mantel coefficient.}
\label{fig:iqr}
\end{figure}

We next attempted to select planets with high potential for having undergone terraforming. Based on the idea that terraformed planets are likely to have similar observable compositions, we start by clustering planets based on only their compositions, at each simulation iteration (see Methods \ref{subsubsec:cluster}). The key question is—of the clusters identified, which ones likely contain terraformed planets? To address this, we selected clusters (without using the ground truth labels of which planets were terraformed) by looking at the effect they have, when removed, on the residual space's Mantel coefficient (Mantel contribution $> 0$; Fig. \ref{fig:mantel_contribution}). We further selected clusters based on their planets' spatial localization (IQR $\leq$ 25.2; Fig. \ref{fig:iqr}).


\begin{figure}[htb!]
\epsscale{1.2}
\plotone{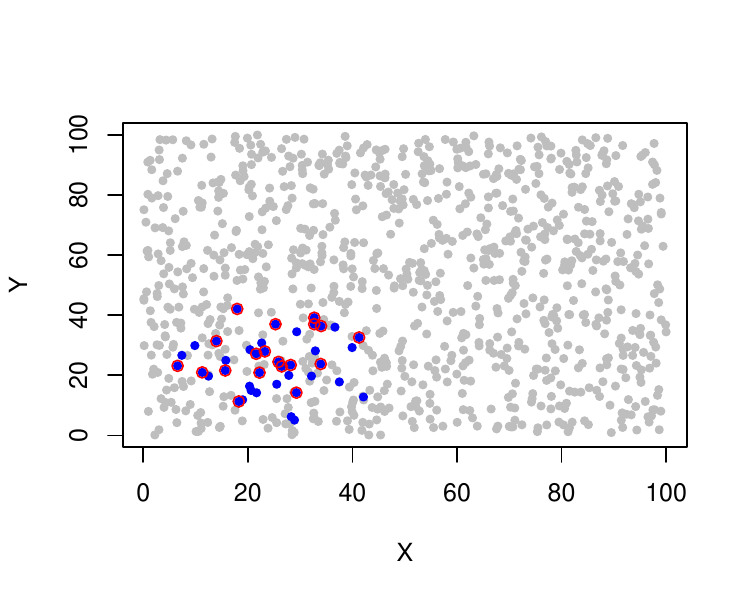}
\caption{
The earliest detected cluster in our simulation, at a terraformed ratio of $0.04$. This is a projection of 3D planet locations in the 2D X-Y plane, and the earliest time step where we detect a cluster of planets meeting our selection criteria. True terraformed planets ($n=40$) have blue fill, while planets detected by our selection method ($n=19$) have a red outline. 
\label{fig:detection_example}}
\end{figure}

\begin{figure}[ht!]
\epsscale{1.2}
\plotone{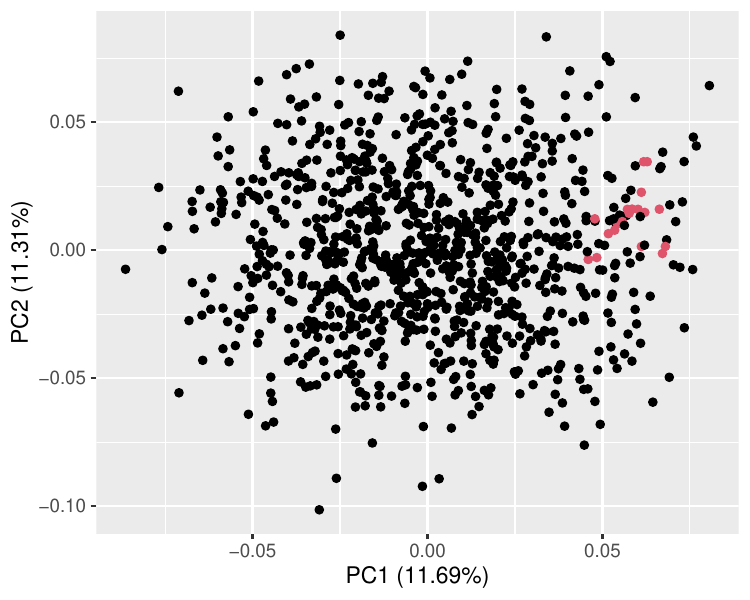}
\caption{
PCA of planetary compositions at a terraformed ratio of $0.04$, the earliest time step where we detect selected clusters (red). PCA is used here only for displaying 10 compositional dimensions on a 2D page, and was not used for clustering.
\label{fig:pca}}
\end{figure}

With both of these selection criteria applied, we detect a total of 247 clusters across all terraformed ratios—the earliest first appearing at a terraformed ratio of 0.04 (Figs. \ref{fig:detection_example}, \ref{fig:pca}). Selected clusters range in size depending on the terraformed ratio, with 1-3 clusters appearing at most simulation iterations between terraformed ratios of 0.04 to 1. Additionally, we find that the Mantel p-value of detected clusters, measured in isolation, is low (Fig.~\ref{fig:clusters_m}).  Changing the parameters used in clustering, or the thresholds used for IQR and MC could identify planet clusters either earlier (by increasing sensitivity) or later (by increasing the IQR threshold). 

\begin{figure*}[htb!]
\epsscale{1.2}
\plotone{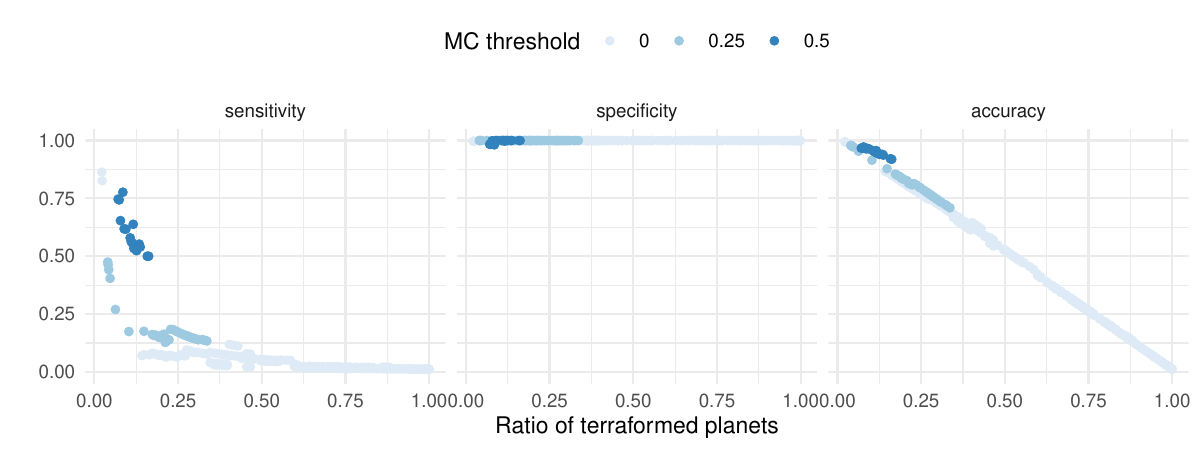}
\caption{
Summary evaluation metrics of sensitivity, specificity, and accuracy for our selected clusters, as a function of terraformed ratio. Color gradient corresponds to subsets of the data meeting MC selection thresholds of 0, 0.25, and 0.5. 
\label{fig:perf_m}}
\end{figure*}

\begin{figure*}[htb!]
\epsscale{1.2}
\plotone{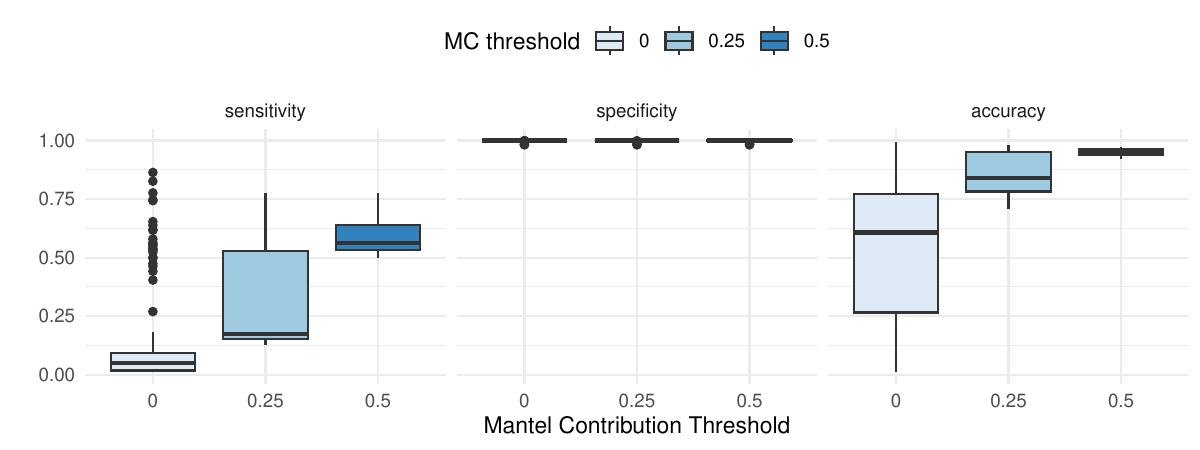}
\caption{Summary evaluation metrics of sensitivity, specificity, and accuracy for our selected clusters, grouped by MC threshold. From left to right, boxes within plots correspond to raising the MC threshold used for selection (left: MC= 0, middle: MC=0.25, right: MC=0.5). 
\label{fig:perf_b}}
\end{figure*}

After selecting clusters, we evaluate if they actually contain terraformed planets (Figs. \ref{fig:perf_m}, \ref{fig:perf_b}). We find that across the full range of terraformed ratios, they have extremely high specificity—close to 1.0—correctly rejecting non-terraformed planets as being non-terraformed. On the other hand, detected cluster sensitivity—a measure of correctly detecting terraformed planets as being terraformed—ranges from 1.0, down to near 0.0. The highest sensitivity clusters appear when fewer planets are terraformed (which is encouraging for the prospects of life detection via our method if we exist at a time when the universe is not yet saturated with life). This could be explained by the fact that throughout the simulation clusters remain small (in terms of number of planets), and the rest of the planets often get classified as noise (this being due to the sensitivity to elbow selection). As a consequence of decreasing sensitivity, accuracy—the proportion of \textit{all} planets correctly classified—also decreases. 

When we raise the MC threshold used for cluster selection (to 0.25 or 0.5), we find that each of these summary statistics improves, reflecting the fact that the more the MC threshold is raised, the more data is excluded from higher terraformed ratios, meaning that the bulk of the data remaining is from early clusters with high sensitivity and high accuracy (Figs. \ref{fig:perf_m}, \ref{fig:perf_b}). 

\section{Summary and Discussion} \label{sec:discussion}

We show how a positive Mantel correlation coefficient with a low p-value can be used to identify when a region of space contains life, under the postulates that life can spread between planetary systems and terraform planets (whether intentionally or not). Going beyond (or below) a population-scale biosignature, we demonstrate how to select, from a population of planets, clusters which have high-likelihood to contain terraformed planets. More importantly, these selected clusters have a\textit{ low-likelihood} to contain \textit{non}-terraformed planets. This is increasingly important as the field of astrobiology reconciles with unavoidable false positives for more traditional biosignatures. In the scenario where our galaxy is already becoming saturated with terraformed life, our detection method becomes less sensitive. Our work has several limitations and space for improvements.

We assumed that the background non-terraformed distribution of planetary compositions is uncorrelated and drawn from a flat random distribution of compositions (a reasonable prior given the high sensitivity of planet formation to initial conditions of the planetary disk, \citealt{tasker_planetary_2020, pacetti_chemical_2022}). In the event that abiotic planets are correlated in position and composition space (perhaps from some undiscovered astrophysical phenomena), then the value of the Mantel correlation coefficient considered to be a statistically significant biosignature would have to increase. It's possible that if the underlying coefficient of non-terraformed planets were too high, the spread of life would actually increase the Mantel p-value, even early in the simulation. If planetary compositions are actually clustered around a few characteristic planet types, we could represent the initial non-terraformed planets as being drawn from a distribution punctuated with narrow Gaussians in composition space, and the results might be different yet. Our results underscore the importance of better understanding the baseline abiotic diversity of planetary populations, including better constraining which features of planet formation best predict planetary characteristics over time. Our model demonstrates that such information could be used to aid life detection in the absence of making progress on a theory of living systems.

Applying our approach to observed astronomical data will also be affected by relative stellar velocities. Naturally this is relevant if we wish to detect potential correlations on the timescale of the evolution of planets. The relative velocity dispersion of stars depends on a variety of factors (such as temperature and age), but in the solar neighborhood generally fall within $20 \leq \sigma_v \leq 40$ km/s \citep{dehnen_local_1998, fatheddin_statistical_2023}. This implies that pairwise distance relationships among stars within a region of scale $L$ become significantly reorganized on a timescale with an upper bound of order $t \sim L/\sigma_v$. Using a value of $L$ corresponding to the solar neighborhood $25 \leq L \leq 100$pc \citep{smart_gaia_2021, golovin_fifth_2023} implies $0.6 \leq t \leq 5$ My. At these timescales, velocity dispersion should dominate the mixing process, when compared to the effects of velocity shear in the Milky Way \citep{binney_galactic_2011}.

Back of the envelope calculations indicate that the potential for panspermia in this time frame may not be so dire with some improvements to spacecraft speed. There are $\sim$ 5200 stars within 25pc of the sun, according to the latest Catalogue of Nearby Stars (CNS5)\citep{golovin_fifth_2023}. This corresponds to a density of $\sim$ 0.08 stars per pc$^3$, or $\approx$ 1 star per 12.5 pc$^3$, and thus an average distance between stars of $\sim$ 3 pc. Assuming as we do in our model that panspermia proceeded exponentially, and in parallel, it would take only $\log_2(40)\sim 5.3$ generations of travel to terraform 40 stars (which is the number we identified in our results as the minimum required for detectability). Six generations corresponds to a minimum of 18 pc (3 pc per generation to the nearest neighboring stars) that have to be traveled sequentially. Benchmarking this against the fastest man made object on trajectory to escape the solar system (the Voyager spacecraft), and the fastest detected interstellar object (comet 3I/ATLAS): terraforming 40 star systems would take $\sim 1$ My years at Voyager speeds ($\sim 15$ km/s  or $1.5 \times 10^{-5}$ pc/yr) \citep{voyager_2024}, or $\sim 250,000$ years at 3I/ATLAS speeds ($\sim 60$ km/s  or $6 \times 10^{-5}$ pc/yr) \citep{team_where_2026} (assuming, as we do in our model, that this also includes the time it takes to terraform). These are long timescales for humans, but within the 5 My upper bound calculated above. Moreover, there are no physical laws preventing the launch of much faster spacecraft (within the speed of light), only technological and economic ones. For example, the Breakthrough Starshot program had planned to create spacecraft that could be accelerated up to speeds of 1000 km/s. If it were possible to engage in panspermia with such spacecraft, the timescale could be reduced to a mere 18,000 years.

In the absolute worst case scenario, the approach we provide would be applicable for stars in similar phase space \citep{lin_statistical_2015}, and of course for life which spreads on average faster than relative stellar velocities—potentially limiting its applicability to life spread via directed panspermia, or via lithopanspermia on exceptionally abundant and fast moving interstellar objects (however, this might cause other challenges in e.g., gravitational capture and impact survival, \citealt{totani_solid_2023}). Overall, stellar and galactic dynamics paint a complicated picture and understanding their impact on our results in more detail is a subject we aim to examine in future work.



Our approach also provides an alternative to specific chemical biosignatures at the level of planets, showing the promise of detecting anomalous features at the population-scale. Critically, these anomalous features can be explained by a model with simple hypotheses about what life does, rather than what life is, and is agnostic to our ignorance of living systems as they might exist outside of Earth. 

By developing a life detection approach that does not require us to assume the existence of specific bio or technosignatures, we can overcome potential issues relevant to the Fermi Paradox. Simple percolation models predict that once a civilization begins the process of galactic expansion, it can proceed to rapidly spread throughout the entire galaxy. Recent work demonstrates that this is contingent on other factors like civilization lifetimes, and it's possible to end up in a steady state where not all planets with the potential to support life will actually contain life, even after settlement fronts have spread across all of galactic space \citep{carroll-nellenback_fermi_2019}. We might expect to have already detected life if we were currently living in such scenarios. One reason this might not be the case is if technology that life is using is too advanced for us to recognize. But another intriguing solution is that life eventually becomes so sustainable that it becomes indistinguishable from a natural non-intelligent or non-biotic planet \citep{schroeder_deepening_2011,likavcan_grass_2025}. The method we proposed here could nevertheless be able to detect interstellar life in either of these scenarios.
 
This discussion is only an introduction to the kinds of questions which could be investigated with this technique. The basic idea of taking key features of life, and understanding how they might manifest on the scale of planets leads us to imagine properties of life that may manifest across scales, like ``planetary phylogenies"—that is, heredities and lineages of planets—and how they could shed light on better understanding living systems generally. Such hypothetical characteristics of planets (and life) could be investigated with our model. 

Perhaps one of the biggest limitations in our model is assuming we can reliably map observable characteristics of a planet to something as comprehensible and malleable as a vector of real numbers. For example, how the chemistry of the atmosphere, biosphere, geosphere, etc. could be reflected via the atmosphere of a planet, and what the mappings are from observed atmospheric properties (the planet's ``phenotype") to the underlying chemistries that put constraints on the compatibility of planets (the planet's ``genotype"), and the kinds of life which can emerge on or coexist with a planet. This would further strengthen the realism of our agnostic approach with minimal assumptions, and in future work we would like to explore these ideas in detail.

While we only walk through a simulation under a single set of parameters here, our goal is to identify a best possible case scenario for applying a technique where we assume life should be able to drive correlations in planetary position-composition space based on hallmarks of life like proliferation and environmental feedback. Our model results show promise that life could be detected at the scale of a population of planets, using information from only $\approx$ 1000 (perhaps fewer) planetary atmospheres, even in the absence of any information about what kinds of planetary environments are most suitable to life, or without knowing anything about the origins of life, or the peculiarities of life's metabolic outputs. We showed how our distributed biosignature can be further refined to detect specific clusters of terraformed planets in this space, even when they only comprise a few percent of all planets. Again, this approach does not require an independent way to ``detect" life, like a smoking gun biosignature atmospheric gas. Instead this approach depends on two key assumptions about what life can do, and derives observable consequences directly from them, providing a statistical approach which can be refined by astronomical surveys, and therein lies its biggest promise.

\begin{acknowledgments}
\section{Acknowledgments}
The authors would like to thank Estelle Janin and Cole Mathis for encouraging and productive conversations, and for feedback on the manuscript.
\end{acknowledgments}
\vspace*{5\baselineskip}




\bibliography{references}{}
\bibliographystyle{aasjournal}


\appendix


\renewcommand{\thefigure}{A\arabic{figure}}
\setcounter{figure}{0}


\section{Appendix} \label{sec:appendix}

\begin{figure*}[ht!]
\plotone{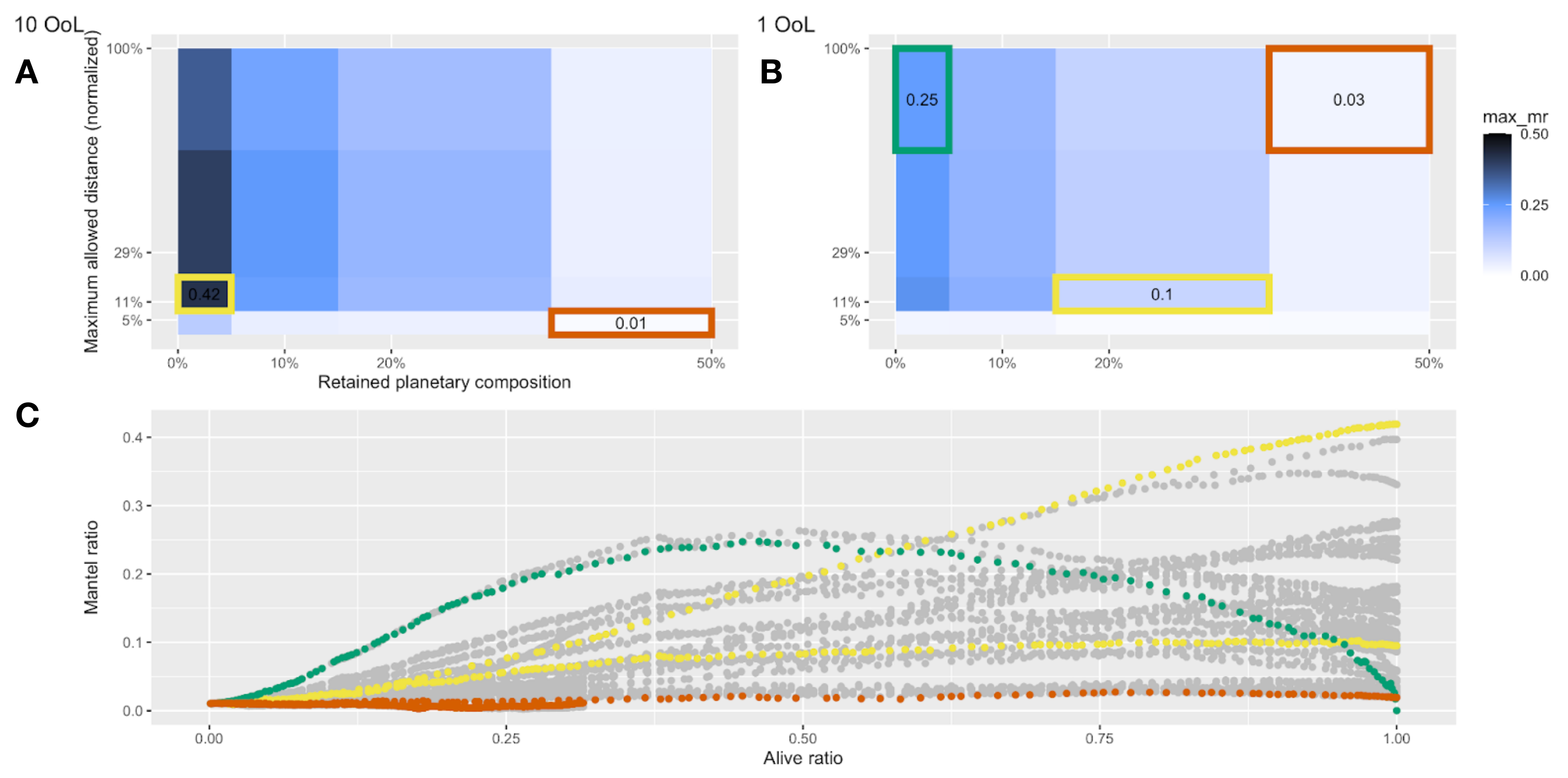}
\caption{
The impact of model parameter changes on the Mantel coefficient. Examples corresponding to ``flat" (orange), ``increasing than decreasing" (green), and ``mostly increasing" (yellow). A and B show heatmaps for the maximum Mantel coefficient reached when varying the the amount of planetary composition retained by the terraformed planet after terraforming (x-axis), and varying the maximum allowed distance for planet target selection (y-axis), for scenarios with 1 OoL (A), and 10 OoL (B). C. The Mantel coefficient across different scenarios as a function of the terraformed planet ratio (x-axis).
\label{fig:param_sweep}}
\end{figure*}

\subsection{The mantel coefficient does not strictly increase with the proliferation of life} \label{subsec:paramsweep}

We investigated how changing the parameters of our model influenced the Mantel coefficient as a function of the ratio of terraformed planets. We find a few characteristic shapes of curves observed: flat, increasing then decreasing, and mostly increasing. 

\subsubsection{Additional Parameters}
\textbf{Mutation. }Depending on the scenario, we include either a 0\% or 10\% chance of mutation for each element of the planet's new composition, meaning that in the 10\% scenario, with our length 10 composition vector, on average, 1 of the 10 elements is mutated during each terraforming event. Mutated elements are chosen from a continuous uniform random distribution of [0, 1], just like the initial compositions were chosen. Mutation is meant to represent the change in the kind of planetary observable characteristics which are compatible with life as life evolves over time. 

\begin{figure}[hbt!]
\plotone{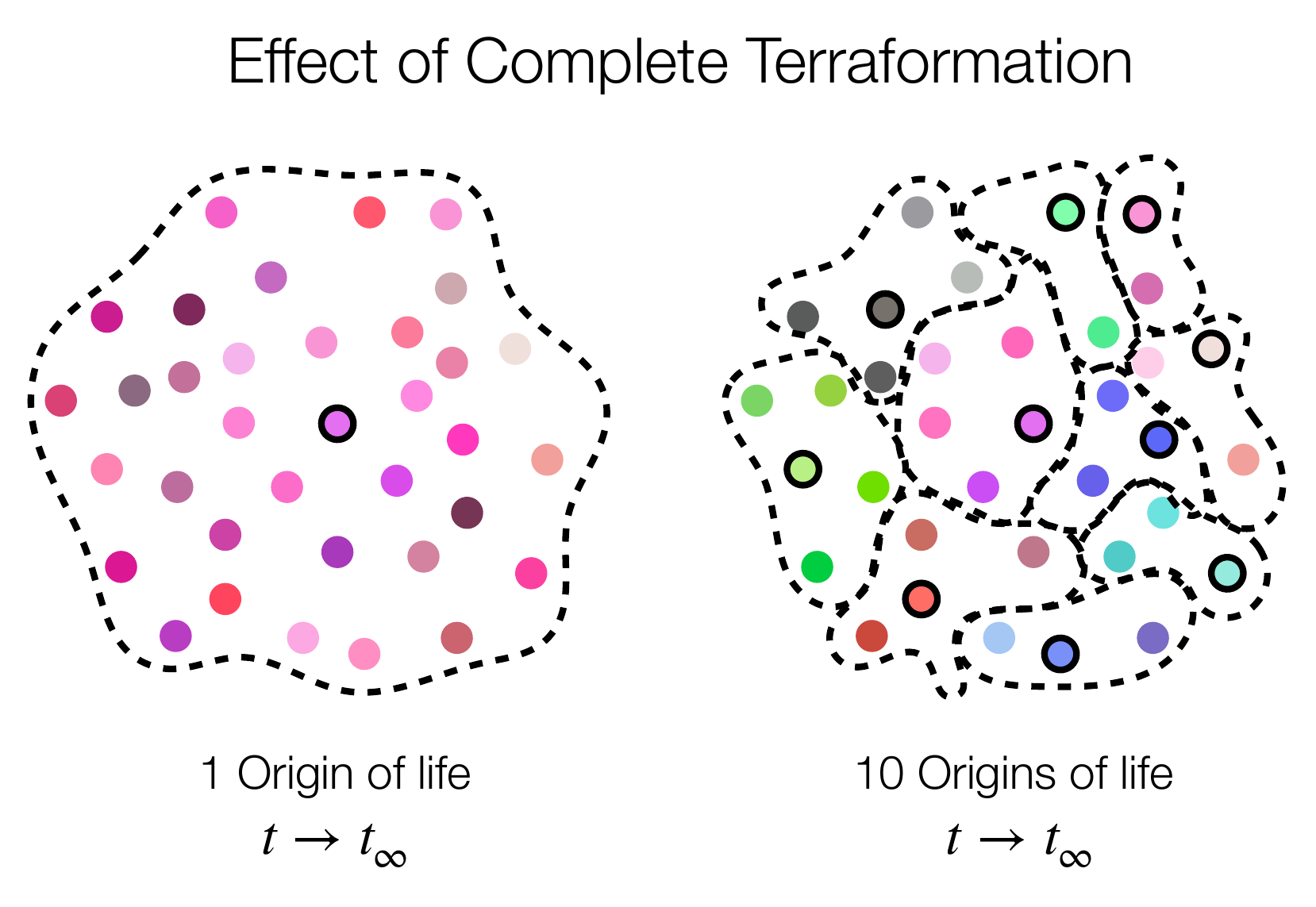}
\caption{
\textbf{Effect of complete terraforming, dependent on number of OoL.} Simulations are initialized with either 1 or 10 origins of life, leading to different dynamics of compositional evolution. }
\label{fig:concept_appendix}
\end{figure}

\textbf{Number of origins of life (OoL).} Simulations are initialized with either 1 or 10 OoL (Fig.~\ref{fig:concept_appendix}).

\textbf{Compositions inherited from pre-terraformed planet}. We vary the number of elements that the post-terraformed planet inherits from the pre-terraformed planet, $ n \in [0,1,2,5]$, with the rest of the composition inherited from the incoming life.

\subsubsection{Flat curve}
Because all simulations are initialized with planets of random compositions, the Mantel coefficient is initially~0. A flat curve shape indicates that the spread of life does not cause a significant increase to the mantel coefficient (Fig.~\ref{fig:param_sweep}). In simulations targeting the most compositionally similar planets within a positional distance cutoff, $min_{comp}(P\leq R_{pos})$, flat curves occur when the distance cutoff is low enough that the terraformed planets run out of new target terraforming candidates (Fig.~\ref{fig:param_sweep}A bottom-most row, orange box).  Flat shapes also occur in scenarios where terraforming maintains at least half of a target planet’s composition (Fig.~\ref{fig:param_sweep}B, rightmost column, orange box).

\subsubsection{Increasing then decreasing curve}
When simulations are initialized with a single OoL, the progression of planet terraforming over time leads to an increasing then decreasing shaped Mantel coefficient curve, where the Mantel coefficient peaks after some of the planets have been terraformed, and then decreases afterwards (Fig. \ref{fig:param_sweep}, data in green). Because the Mantel coefficient is measuring the correlation between positional and compositional distance matrices of planets, this correlation will decrease if either distance matrix becomes too homogeneous. With a single OoL, this is exactly what begins to happen with the compositional distance matrix, with the decline especially pronounced in the scenario with perfect replication (Fig. \ref{fig:param_sweep}B, green box).

\subsubsection{Mostly increasing curve}
The simulations reaching the highest Mantel coefficient are observed when the Mantel coefficient steadily increases as the ratio of terraformed planets increases, peaking near a ratio of 1 (Fig. \ref{fig:param_sweep} data in yellow). This occurs for simulations when the incoming life overwrites most of the terraformed planet’s original composition (Fig. \ref{fig:param_sweep}A, yellow box), and especially when having 10 OoL. In $min_{comp}(P\leq R_{pos})$ simulations, they occur when the distance cutoff is small enough that the simulation doesn’t end before all planets become terraformed, and enable a maximum Mantel coefficient to occur at the point when all planets have been terraformed (Fig. \ref{fig:param_sweep} B, yellow box). The parameters that lead to this behavior appear to be a “tipping point”, where the distance range for panspermia is just barely large enough to allow the simulation to proceed until all planets become terraformed.

\begin{figure}[ht!]
\epsscale{1.2}
\plotone{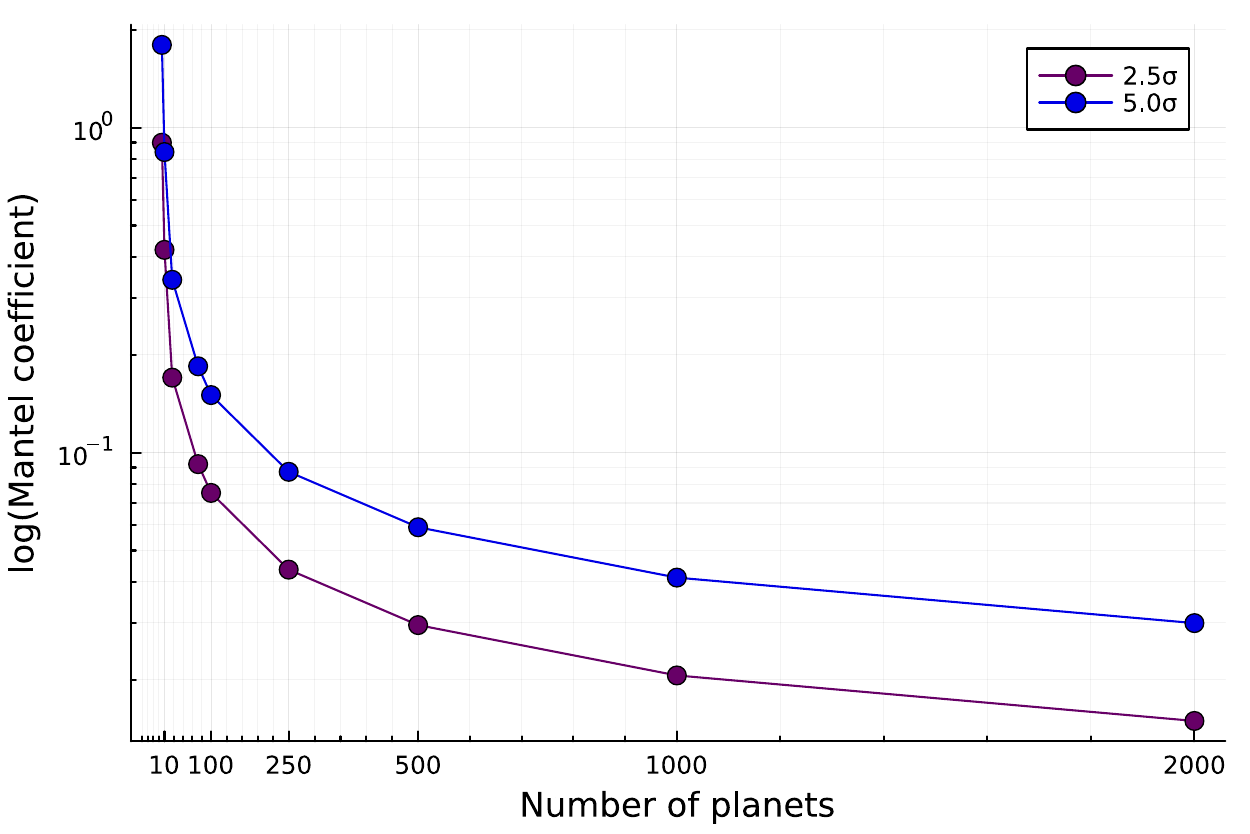}
\caption{A sensitivity analysis of the Mantel coefficient corresponding to a $2.5 \sigma$ and $5 \sigma$ anomaly, as a function of the number of planets observed. Positions and compositions of planets are chosen as in our model (See Methods Sec. \ref{sec:methods}) We find that the sensitivity of the Mantel coefficient to number of planets decreases exponentially, and 1000 planets seems like a reasonable choice to reflect the balance of the challenge of realistically observing planets, with the need for those planets to exhibit potentially small correlations in composition-position space. The exact shape of this plot will vary by model parameters, but is especially dependent on the distribution of planet compositions and positions (e.g., planets being evenly distributed in composition or position space, vs. extremely heterogeneous; not shown)}
\label{fig:test_power}
\end{figure}

\begin{figure}[ht!]
\epsscale{1.2}
\plotone{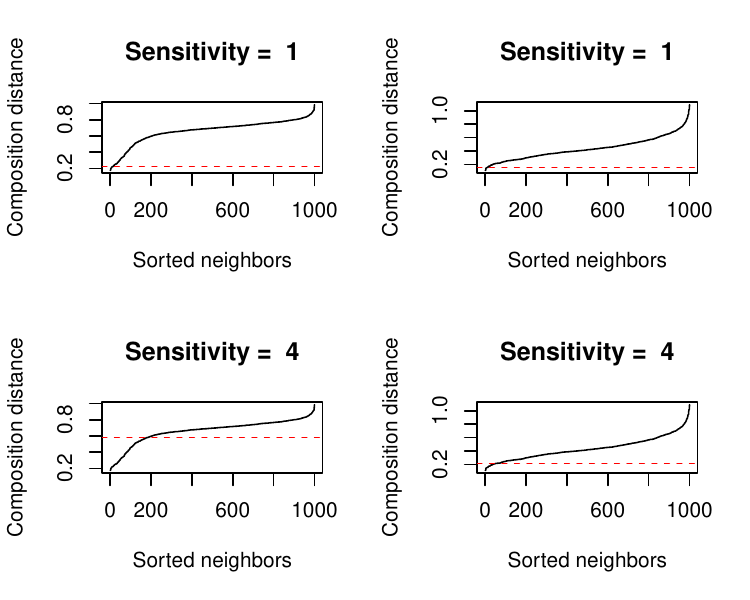}
\caption{Elbow detection using the Kneedle algorithm for t=105 (left) and t = 205 (right) with different values of the sensitivity parameter. The elbow value has a huge influence on the quality of clustering. The curves on the right can be argued to have 3 elbows; 1 steep convex elbow, 1 soft convex elbow, and 1 concave elbow. We are interested in the steepest convex elbow, and therefore choose a sensitivity~=~1 in this paper.
\label{fig:elbow}}
\end{figure}

\begin{figure}[ht!]
\epsscale{1.2}
\plotone{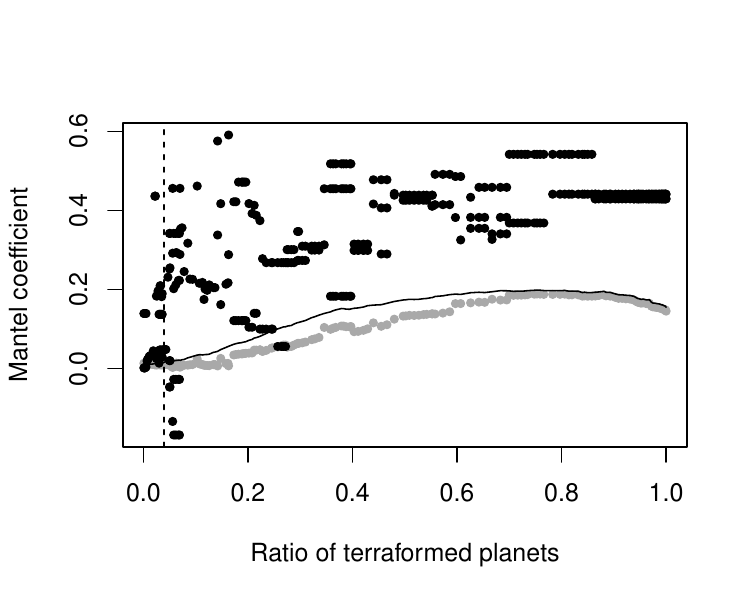}
\caption{Mantel coefficient of each of the clusters as a function of the ratio of terraformed planets in the whole space. In grey are the clusters labeled as ``noise" by the DBSCAN algorithm. The vertical dashed line represents the earliest detection of a cluster of terraformed planets using our proposed approach. The solid curve is the Mantel of the total space.
\label{fig:clusters_m}}
\end{figure}

\begin{figure}[ht!]
\epsscale{1.2}
\plotone{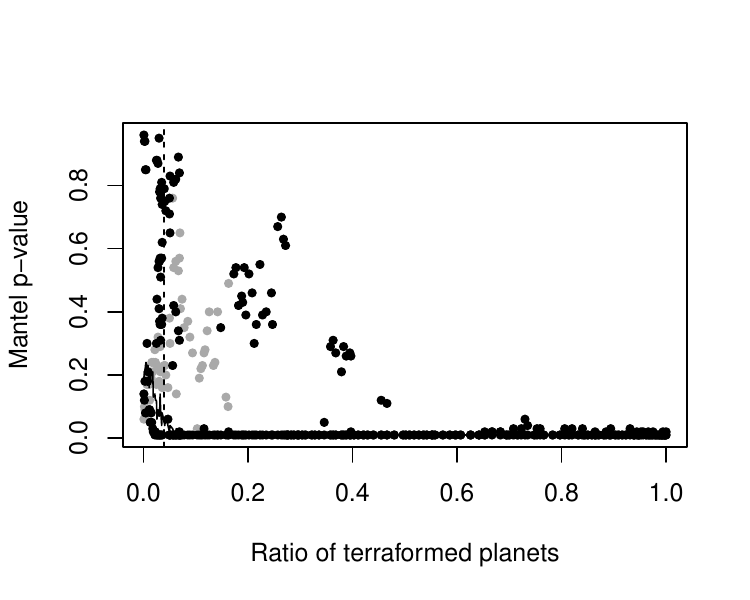}
\caption{P-value of the Mantel coefficient of each of the clusters as a function of the ratio of terraformed planets in the whole space. In grey are the clusters labeled as ``noise" by the DBSCAN algorithm. The vertical dashed line represents the earliest detection of a cluster of terraformed planets using our proposed approach. The solid curve is the p-value of the Mantel coefficient of the total space.
\label{fig:clusters_p}}
\end{figure}

\begin{figure}[ht!]
\epsscale{1.2}
\plotone{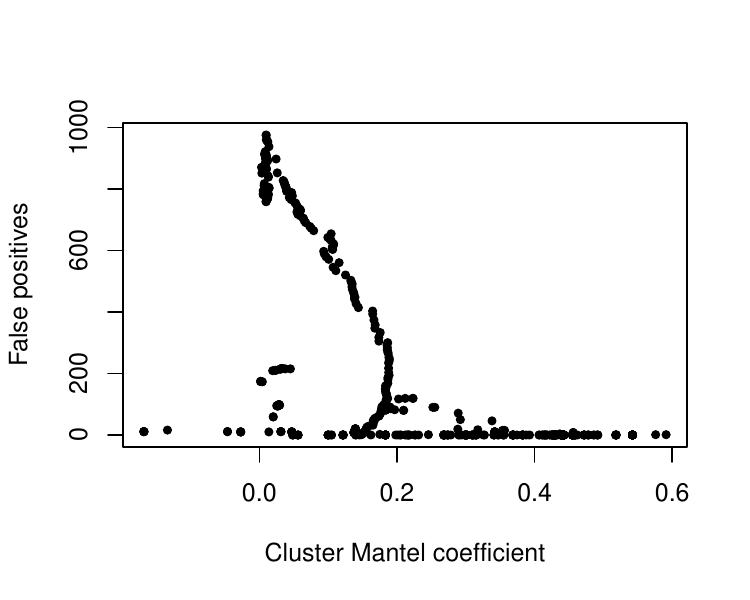}
\caption{Number of false positives in each cluster as a function of the cluster's own Mantel coefficient. While the curve has a fascinating shape, we can see that there is no easy correlation.
\label{fig:corr_m}}
\end{figure}

\begin{figure}[ht!]
\epsscale{1.2}
\plotone{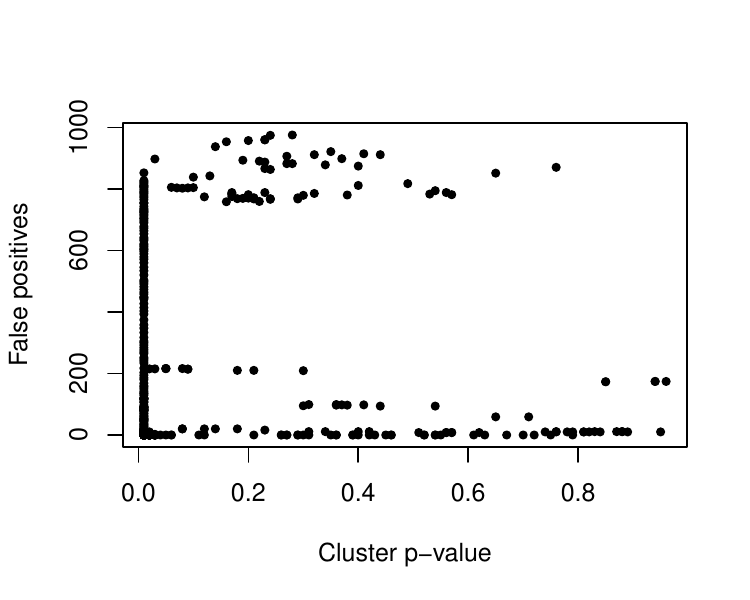}
\caption{Number of false positives in each cluster as a function of the cluster's p-value for its Mantel coefficient. We can see that there is no simple correlation.
\label{fig:corr_p}}
\end{figure}

\begin{figure}[ht!]
\epsscale{1.2}
\plotone{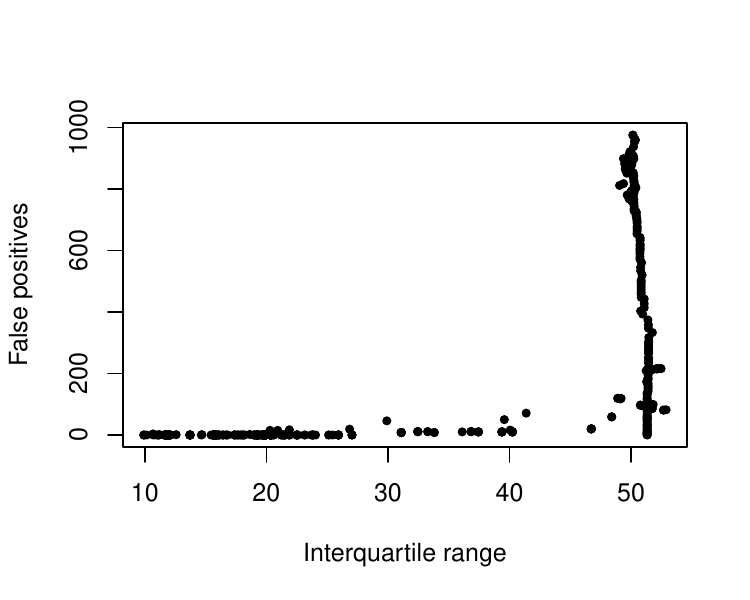}
\caption{Number of false positives in each cluster as a function of the cluster's IQR.
\label{fig:corr_iqr}}
\end{figure}

\begin{figure}[ht!]
\epsscale{1.2}
\plotone{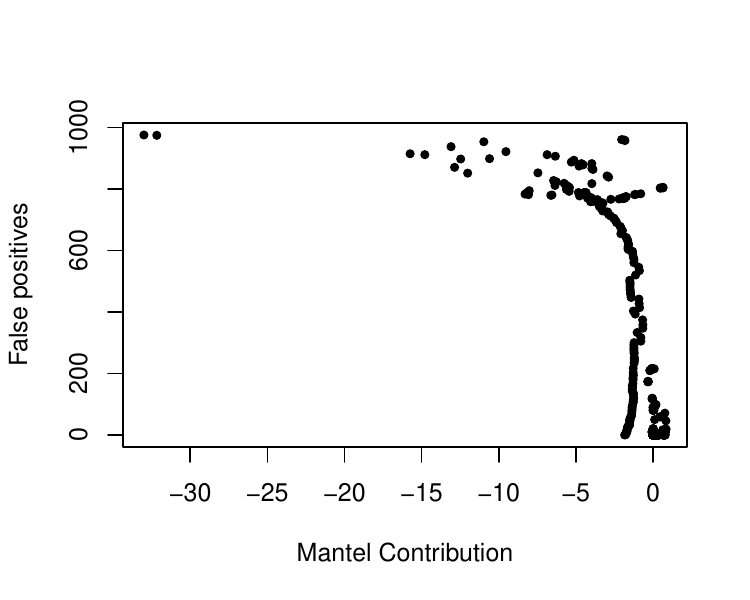}
\caption{Number of false positives in each cluster as a function of the MC. There is a clear divide between the negative and positive MC.
\label{fig:corr_mc}}
\end{figure}

\begin{figure}[ht!]
\epsscale{1.2}
\plotone{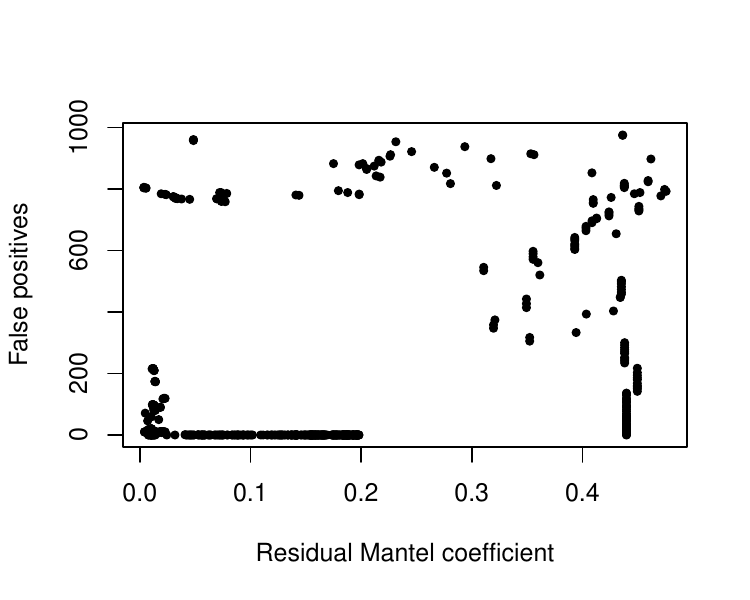}
\caption{Number of false positives in each cluster as a function of the residual Mantel coefficient. We can see that there is no simple correlation.
\label{fig:corr_rm}}
\end{figure}

\begin{figure}[ht!]
\epsscale{1.2}
\plotone{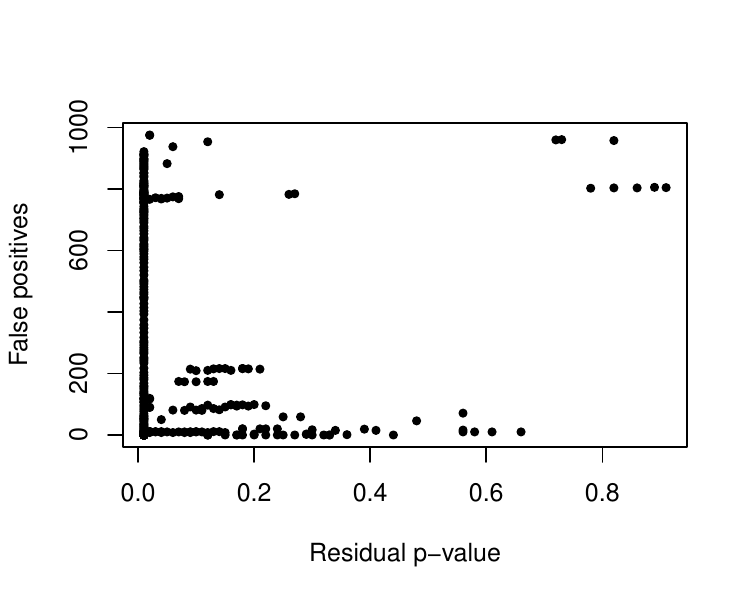}
\caption{Number of false positives in each cluster as a function of the residual p-value. We can see that there is no simple correlation.
\label{fig:corr_rp}}
\end{figure}

\begin{figure*}[ht!]
\plotone{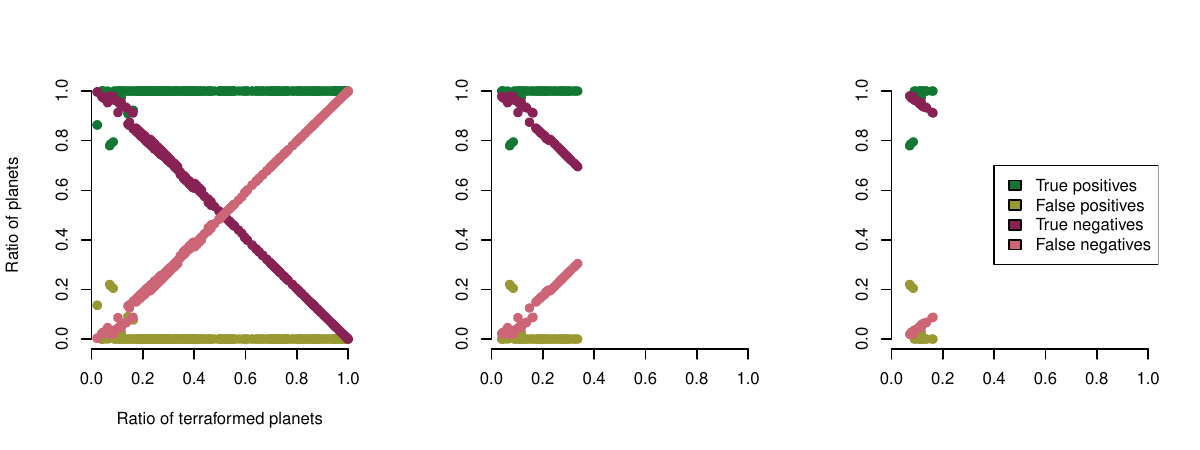}
\caption{Ratios of true positives, false positives, true negatives and false negatives when changing the MC threshold in our selection critera. These results are for clusters below our IQR threshold of $25.2$.
\label{fig:perf_ratios}}
\end{figure*}

\newpage

\end{document}